\documentclass[prb,preprint]{revtex4-1} 

\def\obsA{{\bf A}}
\def\obsB{{\bf B}}
\def\xprime{$x^\prime$}
\def\ctprime{$ct^\prime$}
\def\xct{$x$-$ct$}
\def\xctprime{$x^\prime$-$ct^\prime$}
\renewcommand{\vec}[1]{\mathbf{#1}}

\usepackage{amsfonts} 
\usepackage[left=1in,right=1in, top=1in, bottom=1in]{geometry}
\usepackage{graphicx}
\usepackage{yfonts}
\usepackage{enumerate}
\usepackage[small,compact]{titlesec}
\usepackage{multirow}
\usepackage{amsmath}
\usepackage[caption=false]{subfig}

\begin{document}

\title{A Graphical Introduction to Special Relativity Based on a
  Modern Approach to Minkowski Diagrams}
\author{B.\ Liu}
\email{bliu2@stanford.edu}
\author{T.\ A.\ Perera}
\email{tperera@iwu.edu}
\affiliation{Department of
  Physics, Illinois Wesleyan University, P.\ O.\ Box 2900,
  Bloomington, IL 61702}

\date{\today}

\begin{abstract}
We present a comprehensive introduction to the kinematics 
of special relativity based on Minkowski diagrams and provide a
graphical alternative to each and every topic covered in a standard
introductory sequence. Compared to existing literature on the subject, our introduction of
Minkowski diagrams follows a more structured and contemporary
approach. This work also demonstrates new ways in
which Minkowski diagrams can be used and draws several new insights
from the diagrams constructed.  In this regard, the sections that
stand out are: 1. the derivation of Lorentz transformations (section
\ref{Relax} through \ref{Lorentz}), 2. the discussion of spacetime
(section \ref{Spacetime}), 3. the derivation of velocity addition
rules (section \ref{Velocity}), and 4. the discussion of relativistic
paradoxes (section \ref{Paradoxes}).  Throughout the development,
special attention has been placed on the needs and strengths of
current undergraduate audiences.
\end{abstract}

\maketitle

\section{Introduction}

Most undergraduate physics students encounter Special Relativity and
Quantum Mechanics in their second year through a course on {\em Modern
  Physics.}  As these two topics do not fit particularly well with the
intuition and skills they develop in their first-year courses,
mathematical abstraction seems to be the path forward for many.  If
students are to gain insights and develop intuition in these two
subjects at an early stage, it is important to make available to them
several alternate routes of exploration.  With regard to special
relativity, students welcome the standard undergraduate introduction to Minkowski
spacetime diagrams, guessing that these diagrams will eventually be
helpful for solving quantitative problems.  Unfortunately, this hope
goes unrealized in most Modern Physics texts/courses because the
diagrammatic approach is not developed beyond the descriptive level.
In our experience, this does not stop students from attempting to
adapt spacetime diagrams for quantitative use, often unsuccessfully.
What these students would truly appreciate is the graphical
construction introduced by Minkowski in his famous 1908 lecture on
spacetime diagrams,\cite{Minkowski1908} where the tilted and
stretched \xctprime\ axes of a moving observer are overlaid on the
Cartesian \xct\ grid of a stationary observer (for example, see
Fig.~\ref{relativistic}).  From now on, we will use the term {\em
  Minkowski diagram} to refer to this quantitative graphical
construction, rather than the generic spacetime diagram (or
\xct\ plane) of a single observer.

The main purpose of this work is to showcase the multitude of ways in
which Minkowski diagrams can be used for instruction.  In particular,
we highlight several new ways of using and interpreting Minkowski
diagrams that we have developed.  These together with well established
applications are presented here in one place, as a complete set of
graphical alternatives to all of the standard introductory lessons on
the kinematics of special relativity.  Since standard pedagogy already
makes use of spacetime diagrams in a qualitative sense, extending
their use for quantitative purposes is, we believe, a natural and
expected step.  It is not our hope that the entire sequence laid out
here will be adopted in full.  However, continued access to it will,
we believe, benefit most students and instructors.  We also hope that
this article will put Minkowski diagrams on the same footing as other
(newer) diagrammatic methods by presenting, in one place, a
compilation of its uses, so that instructors may easily gauge the
relative strengths and weaknesses of different methods.

The Minkowski diagram has long been recognized as an effective
quantitative tool in special relativity.\cite{TaylorWheeler1966,
  Shadowitz1968, ShadowitzGRnote} Then, why isn't it used routinely in
introductory treatments as a graphical alternative or to reinforce
standard algebraic methods?  One reason implied in the
literature,\cite{Shadowitz1968} is the need for many geometrical
constructs---triangles and invariant hyperbolae\cite{Minkowski1908,
  Silberstein1914}---and hence, ``busy diagrams'' in obtaining
quantitative results.  After a detailed survey of the literature,
another reason that stands out is the emergence of two other excellent
graphical techniques in the late 1950s to early 1960s, which was also
the time when introductory treatments of special relativity found
their way into second-year undergraduate syllabuses.  In these two
graphical constructions, named Loedel\cite{Loedel1948, Amar1955} and
Brehme\cite{Brehme1962} diagrams, the two observers are treated more
symmetrically than in the Minkowski approach and, as a result, no
stretching of $x$ and $ct$ axes are required for either observer.
Hence, these diagrams were recognized as superior, simpler, and more
appropriate for introductory treatments.  We contend that the ultimate
simplicity of Loedel and Brehme diagrams is achieved through a certain
degree of abstraction and cleverness, which may not be ideal for an
intuition-building introduction to special relativity.  For instance,
identifying/drawing trajectories (worldlines) of the two observers
themselves on a Brehme diagram requires a few steps in
reasoning,\cite{Shadowitz1968, Rekveld1969} unlike with Minkowski
diagrams where these trajectories are obvious.  Furthermore, Loedel
and Brehme diagrams cannot be used to {\em derive} special relativity
from Einstein's two postulates; they are constructed by accepting the
equality of the invariant interval between two inertial
observers.\cite{Rekveld1969}  On the other hand, the Minkowski diagram
can be constructed directly from the postulates, and this is a valid
method of deriving special relativity, as demonstrated in Max Born's
text on relativity.\cite{BornText1962} In addition, the Loedel diagram
and, therefore, the closely-related Brehme diagram can be easily
recognized as special applications of the Minkowski diagram.

Due to the above reasons or perhaps due to the perception that such an
old technique must have already reached its pedagogical potential,
there does not exist a comprehensive introductory treatment of special
relativity, based on Minkowski diagrams, to the best of our knowledge.
Meanwhile detailed introductory sequences based on other graphical
techniques can easily be found in textbooks\cite{TaylorWheeler1966,
  Shadowitz1968} and pedagogy-oriented publications including this
journal.\cite{AJPintros}  The closest parallels to these treatments,
that utilize Minkowski diagrams, can be found on the worldwide
web,\cite{WebNote} not in print journals.  On the other hand, {\em
  specific} applications of Minkowski diagrams have appeared in the
past literature\cite{PastLitNote} as well as in more recent journal
articles.\cite{Reynolds1990Cook1991} Unfortunately, authors of
recent graphical treatments seem to be unaware of the connection
between their methods and the original work of Minkowski.

Over the past 2 years, we have assembled a complete and original
lesson plan for introducing special relativity at the second-year
level purely through Minkowski diagrams.  The complete cannon of
introductory topics including the derivation of Lorentz
transformations, length contraction, time dilation, velocity addition,
Doppler shift, and an exposition of well know paradoxes are covered.
In applying Minkowski diagrams to these topics, we have tried to use
modern arguments that would be most transparent to current
second-year students, given their usual preparation and experience at
this stage.  In particular, we have avoided excessive use of
geometrical constructs including invariant hyperbolae.  Instead, we
have pursued a unique approach where where the \xctprime\ grid pattern
of a moving observer is examined using a mix of geometry and algebra.
We have also tried to develop methods that are useful for solving
standard textbook problems in special relativity.  These goals were
achievable mainly because the work was carried out as a
student-faculty collaboration.  The result of this work, which we
present here, is a streamlined introductory sequence made up of
succinct individual lessons that are often quite different from
arguments/derivations we have seen in the previous literature or on
the web.

We start in section \ref{Galilean} with Galilean transformations and
an introduction to the graphical approach used here.  In section
\ref{Derivation}, using Einstein's postulates and the previously
developed graphical ideas, work toward Lorentz transformations and the
other kinematical results of special relativity.  In section
\ref{Results}, we demonstrate the usefulness of Minkowski diagrams in
deriving well know results and solving typical textbook problems.  In
section \ref{Paradoxes}, we demonstrate the use of this method in
unraveling several well known paradoxes of special relativity.  In
this work, we will not cover the standard topics discussed in the
context of generic spacetime diagrams, such as worldlines and light
cones.  Those topics can be introduced prior to or in parallel with
the development presented here.

\section{Minkowski diagrams and Galilean transformations}
\label{Galilean}

\begin{figure}
\centering
\includegraphics[width=3.5in]{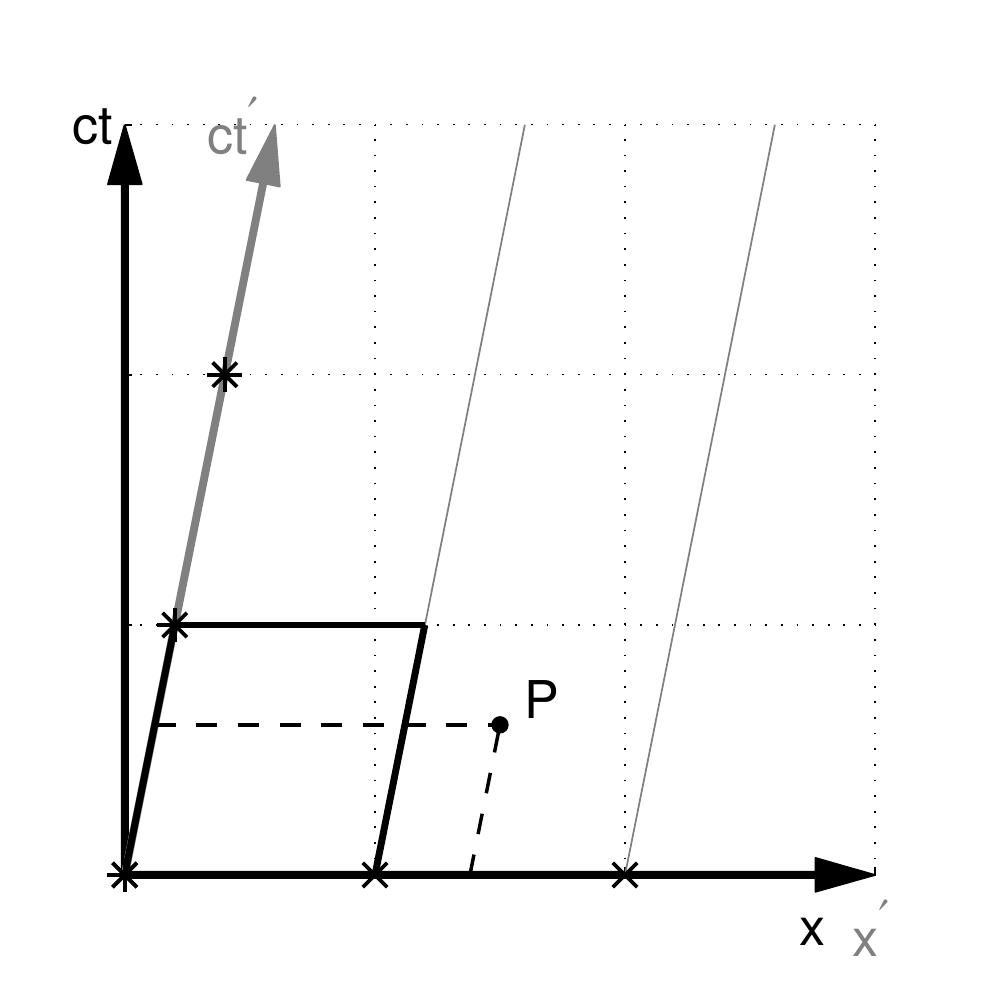}
\caption{The overlaid spacetime grids of observers \obsA\ and \obsB.
  The crosses represent clocks in \obsB's frame separated by a unit
  length.  The stars on the \ctprime\ axis represent the ticks of the
  clock placed at $x^\prime = 0.$ The parallelogram in dark outline is
  the unit cell of \obsB's grid.  The dashed lines indicate how event
  P can be projected onto the \xctprime\ grid.}
\label{galilean}
\end{figure}

We introduce Minkowski diagrams in the context of Galilean
transformations so that methods can be introduced independently from
the surprises of relativity.  Initially, spacetime diagrams may be
referred to as ``time vs. position graphs,'' as the unification of
space and time comes later.  On these position-time graphs
(e.g.\ Fig.~\ref{galilean}), one spatial dimension ($x$) and the $ct$
axis will be displayed, following convention.  Thus, the trajectories
(worldlines) of light pulses have slopes of $\pm 1.$

On Fig.~\ref{galilean}, we first note the \xct\ axes and square
(dotted-line) grid of observer \obsA, who is stationary relative to
the page.  We regard \obsA's lines of constant $x$ (vertical dotted
lines) as the worldlines of synchronized clocks spaced apart by unit
length intervals along his $x$ axis.  Along these worldlines, we mark
a set of points corresponding to the ticks of each clock.  The lines
of constant $ct$ (horizontal dotted lines) are the lines connecting
these clock ticks.  Next, we overlay on this graph the \xctprime\ grid
of observer \obsB, who is moving with speed $v$ in the $+x$ direction
of observer \obsA.  We do so by following the ticks and worldlines of
equally-spaced synchronized clocks in observer \obsB's reference
frame.  The positions of these clocks at $t = 0$ are marked by the
crosses of Fig.~\ref{galilean}.  For convenience, the origins of the
two grids intersect at $t = t^\prime = 0$ and the spatial axes of the
two coordinate systems are aligned with each other.  We can construct
observer \obsB's position-time grid by adopting two ``postulates''
from {\em day to day experience:} {\bf (1)} the size of a unit ruler
does not change due to one's motion, and {\bf (2)} all the clocks of a
moving observer remain synchronized with one another and with the
clocks of a stationary observer.  It follows that the titled (light)
solid lines of Fig.~\ref{galilean}, with slope $c/v,$ are lines of
constant \xprime\ while the horizontal dotted lines serve as lines of
constant \ctprime\ as well.  This type of diagram, where the
position-time grids of both observers are overlaid, is what we will
refer to as a {\em Minkowski diagram.}  In section \ref{Derivation},
we will replace the familiar postulates {\bf (1)} and {\bf (2)} with
Einstein's postulates to arrive at the relativistically correct form
of the Minkowski diagram.

Now, given the $x$ and $ct$ coordinates of an event, such as event P
of Fig.~\ref{galilean}, one can project it onto the \xctprime\ grid as
indicated by the dashed lines.  Using simple geometry and the fact
that there is no motion along the $y$ and $z$ axes, it is easy to show
that
\begin{eqnarray}
x^\prime = x - vt, ~~~ y^\prime = y, ~~~ z^\prime = z, ~~~ t^\prime = t.
\label{galilean_transformations}
\end{eqnarray}
Eqs.~\ref{galilean_transformations} are the Galilean transformations,
which we have derived using postulates {\bf (1)} and {\bf (2)} above.
Before starting on relativity, it is important to point out some key
results that follow from Fig.~\ref{galilean}.  An object that moves
with a constant velocity according to observer \obsA\ would be
represented by a straight line in Fig.~\ref{galilean}.  This worldline
would have a constant rise over run (slope) on the \xctprime\ grid as
well.  Therefore, the object has a constant velocity according to
observer \obsB\ as well.  However, the \xprime\ velocity will {\em
  always} differ from the $x$ velocity by $v,$ the relative velocity
between \obsA\ and \obsB.

\section{Graphical derivation of the kinematics of special relativity}
\label{Derivation}

We begin with Einstein's two postulates.  A definition/explanation of
inertial observers, as those for whom the laws of physics assume
their familiar and simple forms, should precede this.

\noindent
{\bf 1.} If an observer moves with constant velocity relative to
an inertial observer, he/she is an inertial observer as well.

\noindent
{\bf 2.} All inertial observers will obtain/measure the same numerical
value $c$ for the speed of light.

We note that under these two postulates, it is still possible for each
inertial observer to set up a system of synchronized clocks as before.
Many introductory texts on relativity describe in detail how a 3-d
jungle gym of clocks can be synchronized, most often using light
signals, in a world where the above postulates are
true.\cite{TaylorWheeler1966} Thus, given that \obsA\ is inertial, his
position-time grid can still be represented by the dotted lines of
Fig.~\ref{galilean}.  Now, according to postulate 1 above, \obsB\ is
also inertial, as he moves with constant velocity relative
to \obsA.  However, it is obvious that observer \obsB's coordinate
grid can no longer be represented as in Fig.~\ref{galilean} because,
according to that representation, the speed of a light pulse moving in
the positive (negative) $x$ direction of observer \obsA\ would be $c -
v ~ (c + v)$ for \obsB.  Our goal is to find the correct
representation of \obsB's grid.  We will start by asking which aspects
of Fig.~\ref{galilean} we should keep and which aspect we need to
change.

\subsection{Relaxation of Galilean assumptions}
\label{Relax}

\noindent
{\bf 1.}  Observer \obsB's \ctprime\ axis and other lines of constant
\xprime\ should continue to be straight lines with slope $c/v,$
because they are the worldlines of clocks that move along with
observer \obsB\ at speed $v$ relative to observer \obsA.

\begin{figure}
\centering
\includegraphics[width=3.5in]{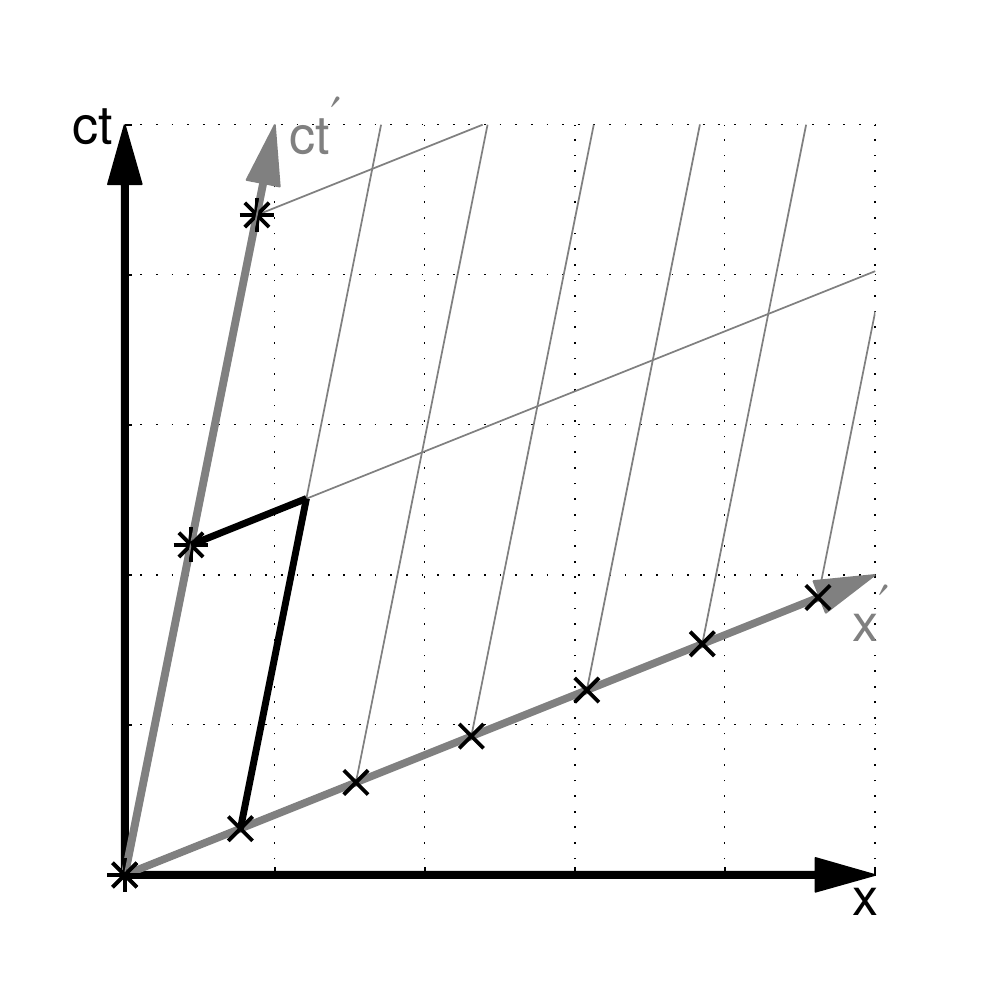}
\caption{Here, we capture our present understanding of what \obsB's
  $x^\prime$-$c t^\prime$ coordinate system may look like.  The
  spacing between crosses (unit length markings), the spacing between
  stars (unit time markings), and the angle of the $x^\prime$ axis
  have been chosen arbitrarily here, as we have not determined those
  parameters thus far.  An important result is that \obsB's coordinate
  grid can be completely characterized by the repeating unit-cell
  parallelogram highlighted here.}
\label{freedoms}
\end{figure}

\noindent
{\bf 2.}  Next, we ask if the stars along the \ctprime\ axis of
Fig.~\ref{galilean}, which are the ticks of \obsB's clock at unit time
intervals, could be spaced differently than they are.  It is easy to
appreciate that such a change would cause \obsB's speed of light
measurements to yield values different from $c \pm v.$ Physically, we
are asking whether we (and observer \obsA) might observe \obsB's
clocks to be ticking at a different rate from \obsA's clocks due to
\obsB's motion, even though all clocks were manufactured identically.
Although we will not answer this question definitively just yet, let
us keep the option to change the spacing between stars on the
$ct^\prime$ axis.  We have represented this freedom in
Fig.~\ref{freedoms}.  However, we require that the stars are {\em
  equally} spaced along the ($ct^\prime$) axis.  Physically, this is
equivalent to requiring that the relationship between \obsA's clocks
and \obsB's clocks depends solely on their relative motion and not on
the specific time on an observer's clock or the absolute distance
between observers.

\noindent
{\bf 3.} Next, we turn our attention to the $x^\prime$ axis.  Our
guiding principle here will be that all observers identified as being
inertial by \obsA\ must also appear inertial to \obsB.  Thus, all
straight lines in \obsA's coordinate system should have a constant
slope (rise over run) in \obsB's coordinates as well.  This constrains
the $x^\prime$ axis to be represented by a straight line and his unit
length markings along $x^\prime$ (crosses) to be equally spaced.  For
the moment, we acknowledge the possibility that the $x^\prime$ axis
may not be parallel to that of observer \obsA\ and that the unit
length marks on it may be spaced differently from the unit length
marks on \obsA's $x$ axis.  We will elaborate on the physical
significance of these potential differences later.

Fig.~\ref{freedoms} encapsulates our present reasoning on what \obsB's
position-time grid may look like.  An important result is that \obsB's
position-time grid can be completely characterized by the repeating
parallelogram indicated in Fig.~\ref{freedoms}.  Its sides represent
\obsB's unit length and time intervals.  Therefore, we concentrate on
this ``unit cell'' from now on.

\subsection{The shape of observer \obsB's unit Cell}

\begin{figure}
\centering
\subfloat[]{
\includegraphics[width=0.4\textwidth]{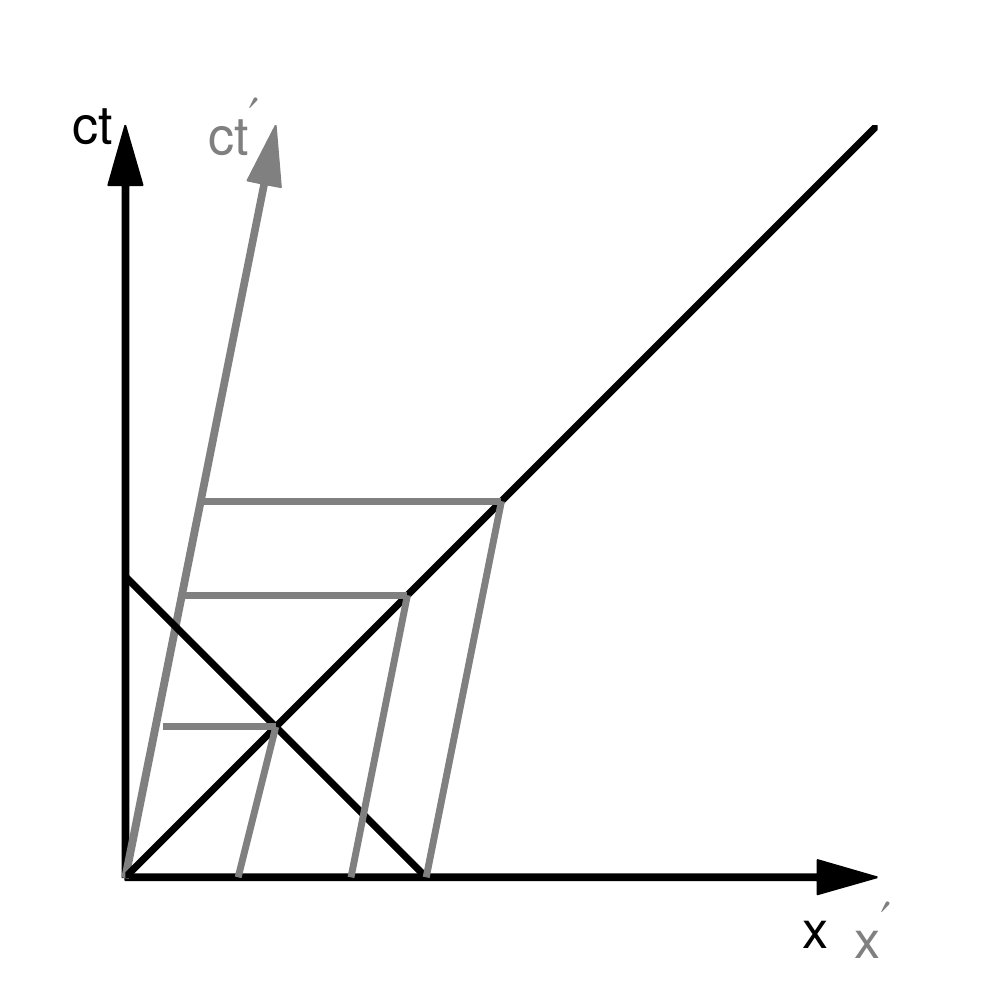}
\label{bad_cells}
}
\subfloat[]{
\includegraphics[width=0.4\textwidth]{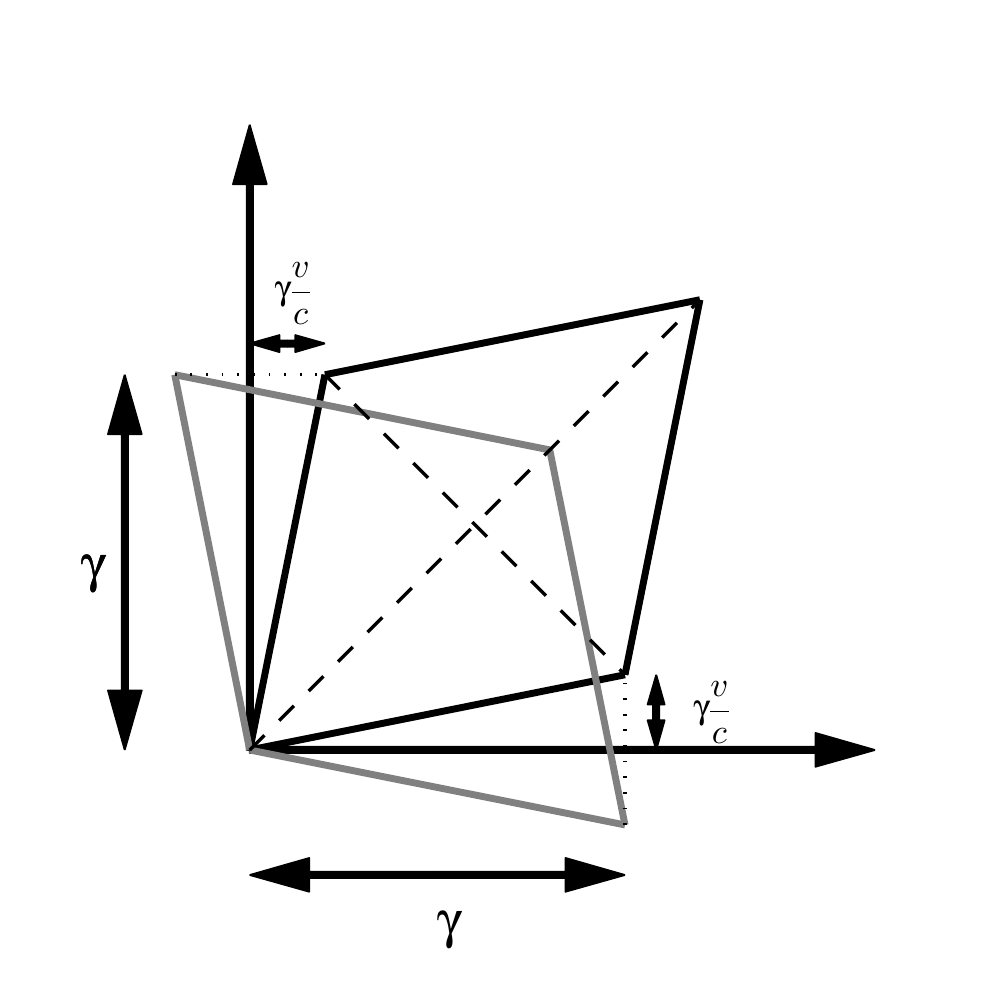}
\label{good_cells}
}
\caption{(a) Several choices of \obsB's unit cell that would yield
  $c,$ (or a slope of 1) in \obsB's frame for a light pulse moving to
  the right.  For these unit cells, the $x^\prime$ axis is horizontal
  as in Galilean transformations.  (b) The parallelogram outlined by
  the dark solid line is the correct representation of \obsB's
  unit cell.  The parallelogram outlined by the light solid line is
  the representation of \obsA's unit cell by \obsB.}
\label{fig3}
\end{figure}

So far, the undetermined properties of the unit cell are the lengths
of its $x^\prime$ and $ct^\prime$ sides and the slope (or angle) of
the $x^\prime$ side.  We can easily utilize these degrees of freedom
so that the speed of a particular light pulse will be measured as $c$
(not $c - v,$ for example) by observer \obsB.  For instance, consider
a light pulse traveling in the $+x$ direction.  According to observer
\obsA, its worldline has a slope of 1 and can be represented by the
long dark solid line of Fig.~\ref{bad_cells}.  This figure also shows
several options for \obsB's unit cell that will yield a slope of 1 in
\obsB's spacetime diagram.  In these trial unit cells, the $x^\prime$
side has been kept horizontal.  The problem with the trial unit cells
of Fig.~\ref{bad_cells} is that they do not yield the correct speed
for a light pulse traveling in the $-x$ direction, represented in
Fig.~\ref{bad_cells} by the dark solid line with slope -1.  The
problem is that this line does not connect two opposite vertices of
the trial unit cells.  Therefore, if we were to construct \obsB's
position-time grid from these unit cells, the worldlines of left-ward
moving light pulses would not have a slope of -1.  The requirement
that both light pulses must travel a unit length per unit time in
\obsB's frame results in the following geometrical constraint on
\obsB's unit cell:

{\em The two lines that connect opposite vertices must have slopes of
  +1 and -1.}

After some exploration, students will realize that the correct unit
cell must look like the parallelogram in dark outline in
Fig.~\ref{good_cells}.  The dashed lines are the lines of slope $\pm
1$ that connect opposite vertices of the unit cell.  It is easy to
show that the four right triangles separated by the dashed lines are
identical.  Therefore, we find that (1) the $x^\prime$ and $ct^\prime$
sides of the unit cell have the same length, and (2) the slope
of the $x^\prime$ side is $v/c,$ the inverse of the slope of the
$ct^\prime$ side.

\subsection{The relativity of simultaneity}

At this, point we recommend pointing out the physical significance of
observer \obsB's \xprime\ axis being tilted relative to \obsA's $x$
axis.  The $x$ axis identifies events that are simultaneous with $t =
0$ in observer \obsA's reference frame.  Similarly, the \xprime\ axis
identifies events that are simultaneous with $t^\prime = 0$ in \obsB's
reference frame.  Earlier, we required that the two coordinate origins
pass though each other at $t = t^\prime = 0.$ From the shape of
observer \obsB's unit cell (Fig.~\ref{good_cells}), we can clearly see
that \obsA\ and \obsB\ identify {\em different sets of events} as
being simultaneous with that event, and hence that simultaneity is a not an
absolute concept, but rather depends on one's reference frame.  The
surprising implications of this truth will be highlighted in sections
\ref{Results} and \ref{Paradoxes}.

\subsection{Lorentz transformations}
\label{Lorentz}

Having established the shape observer \obsB's unit cell, the next step
is to find its size.  This is tantamount to finding the value of the
``scaling constant'' $\gamma$ in Fig.~\ref{good_cells}.\cite{GammaGeom} To
do so, we must use the concept of reciprocity between observers
\obsA\ and \obsB.  The particular method that we follow, makes use
of some very basic linear algebra.  We recommend this approach because
it (1) automatically yields Lorentz transformations, and (2) fits
better with modern pedagogy in terms of viewing Lorentz
transformations themselves as mathematical objects, or operators.
Most second year students are familiar with matrices and matrix sums.
In addition, they have or will soon encounter linear algebra in their
course work and this derivation will serve as a prelude or
reinforcement of those concepts.  In Appendix I, we provide a more
traditional geometrical derivation of the value of $\gamma.$

We start by considering observer \obsB's point of view.  On his
position-time graph, the $x^\prime$ and $ct^\prime$ axes are
orthogonal and the unit interval markings (tick marks) on both these
axes are spaced precisely one unit apart.  In moving to observer
\obsA's representation of this coordinate system, these two axes tilt
and the space between tick marks may change, as far as we know.  In
Fig.~\ref{good_cells}, the parallelogram in dark outline represents
this transformation.  We note, however, that (a) the \xprime\ and
\ctprime\ axes are still straight lines, and (b) the space between
tick marks on these axes, even if different from 1, remains a
constant.  As properties (a) and (b) are synonymous with {\em linear
transformations,} we can now find the matrix representation of this
transformation.  We do this by following how unit vectors in \obsB's
coordinate system get represented in \obsA's coordinate system.  Let
$\begin{pmatrix} 1 \\ 0 \end{pmatrix}$ and $\begin{pmatrix} 0
  \\ 1 \end{pmatrix}$ be unit vectors along the \xprime\ and \ctprime\
axes respectively, as represented on observer \obsB's position-time
graph.  According to Fig.~\ref{good_cells}, the move to \obsA's
coordinate system has the following effect:
\begin{eqnarray*}
\begin{pmatrix} 1 \\ 0 \end{pmatrix} & \longrightarrow
& \begin{pmatrix} \gamma \\ \gamma v/c \end{pmatrix}\\
\\
\begin{pmatrix} 0 \\ 1 \end{pmatrix} & \longrightarrow
& \begin{pmatrix} \gamma v/c \\ \gamma \end{pmatrix}.\\
\end{eqnarray*}
Therefore, the matrix that performs the this transformation must be
\begin{eqnarray}
L_{{\bf B} \to {\bf A}} = \gamma \begin{pmatrix} 1 & v/c \\ v/c &
1 \end{pmatrix}.
\end{eqnarray}
Thus a general event represented by the vector $\begin{pmatrix} x^\prime
  \\ ct^\prime \end{pmatrix}$ in \obsB's reference may be transformed into
\obsA's frame as
\begin{eqnarray}
\begin{pmatrix} x \\ ct \end{pmatrix} 
= \gamma \begin{pmatrix} 1 & v/c \\ v/c & 1 \end{pmatrix}
\begin{pmatrix} x^\prime \\ ct^\prime \end{pmatrix}.
\end{eqnarray}

A move from \obsA's coordinate system to \obsB's proceeds in a
perfectly symmetric manner.  The parallelogram with the light outline
in Fig.~\ref{good_cells} shows how \obsA's unit cell would be
represented in \obsB's position-time graph.  This parallelogram
differs from the first one only in terms of its sides having negative
slopes, due to observer \obsA\ moving in the $-x^\prime$ direction of
observer \obsB.  Therefore, using similar reasoning as above, we find
that the matrix for transforming events from \obsA's coordinates to
\obsB's is
\begin{eqnarray}
L_{{\bf A} \to {\bf B}} = \gamma \begin{pmatrix} 1 & -v/c \\ -v/c &
1 \end{pmatrix},
\end{eqnarray}
where $\gamma$ stands for the very same scaling constant as before.
Using $L_{{\bf A} \to {\bf B}},$ we can conveniently convert an
event's $\begin{pmatrix} x \\ ct \end{pmatrix}$ coordinates to
$\begin{pmatrix} x^\prime \\ ct^\prime \end{pmatrix}$ coordinates.  Of
course, we need to know the value of $\gamma$ to fully define the
matrix operators $L_{{\bf A} \to {\bf B}}$ and $L_{{\bf B} \to {\bf
    A}}.$ We can find $\gamma$ by requiring that a vector remains
unchanged if the two transformations are performed in sequence on an
event, or
\begin{equation}
L_{{\bf B} \to {\bf A}} L_{{\bf A} \to {\bf B}} = \gamma \begin{pmatrix} 1 & v/c \\ v/c &
1 \end{pmatrix} \gamma \begin{pmatrix} 1 & -v/c \\ -v/c &
1 \end{pmatrix} = \begin{pmatrix} 1 &
0 \\ 0 & 1 \end{pmatrix}
\end{equation}
Carrying out the matrix sum, we finally find that
\begin{eqnarray}
\gamma = {1 \over \sqrt{1 - v^2/c^2} }.
\end{eqnarray}
In the process, we have derived Lorentz transformations in the form of
Eqs. 2 and 3.

Actually, a complete Lorentz transformation must be represented by a
4-dimensional matrix and we still need to show that its second and
third rows/columns, corresponding to the $y$ and $z$ transformations,
are trivial.  First, it is easy to argue that the \xprime\ and
\ctprime\ axes cannot have projections along the $y$ and $z$ axes due
to the complete symmetry of the situation with regard to those
dimensions.  Thus, the second and third rows can only have diagonal
elements.  The requirement that $L_{{\bf B} \to {\bf A}} L_{{\bf A}
  \to {\bf B}} = 1,$ allows us to conclude that the diagonal elements
are 1.  Thus, the complete Lorentz transformation will have the form
\begin{eqnarray}
L_{{\bf A} \to {\bf B}} = \begin{pmatrix} 
\gamma & 0 & 0 & -\gamma v/c \\
0 & 1 & 0 & 0 \\
0 & 0 & 1 & 0 \\
-\gamma v/c & 0 & 0 & \gamma \\
\end{pmatrix},
\label{lorentz_matrix}
\end{eqnarray}
Written out as individual equations, the final form of the Lorentz
transformation is
\begin{eqnarray}
x^\prime = \gamma (x - vt), ~~~~~~~~
y^\prime = y, ~~~~~~~~
z^\prime = z, ~~~~~~~~
t^\prime = \gamma(t - v x / c^2).
\label{lorentz_transformations}
\end{eqnarray}
The opposite transformation (from \obsB's frame to \obsA's) looks
identical except for a switch in the signs preceding $v.$

\subsection{Use and construction of relativistically correct
Minkowski diagrams}
\label{Recipes}

\begin{figure}
\centering
\includegraphics[width=3.5in]{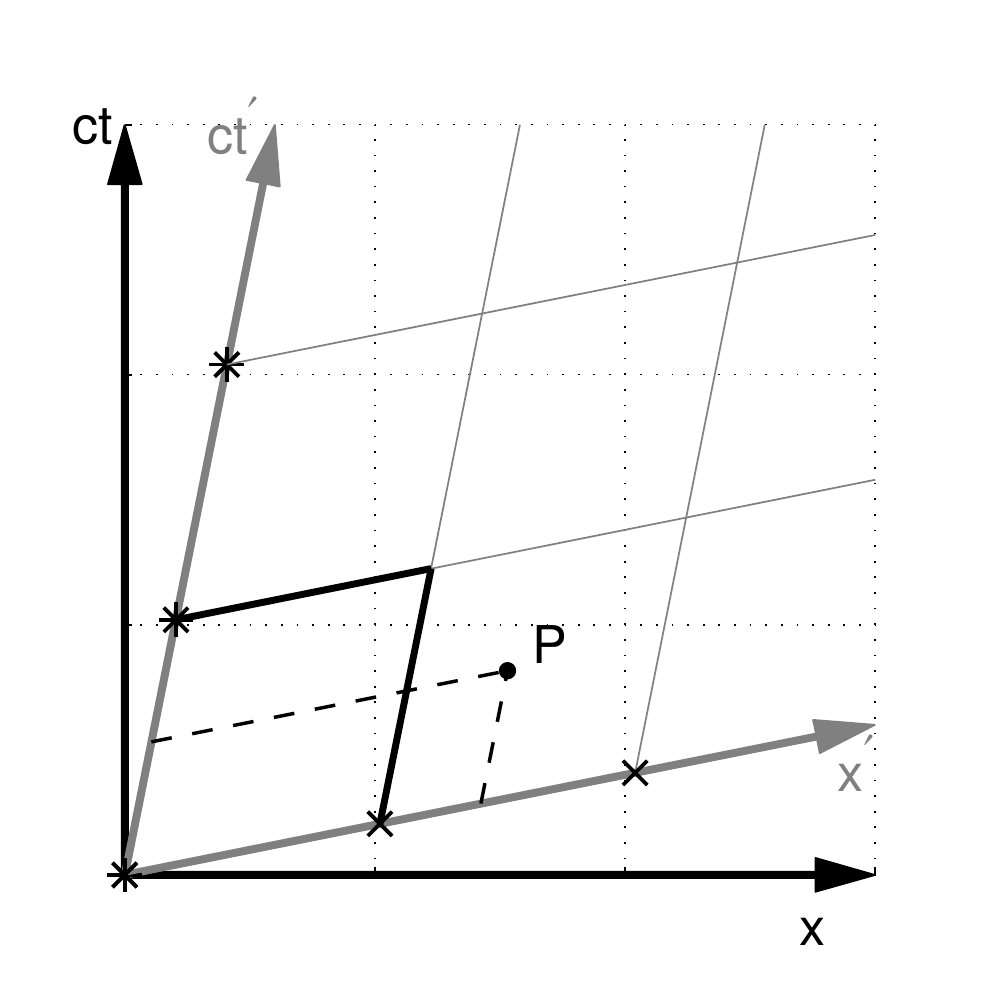}
\caption{Relativistically correct Minkowski diagram showing the
  position-time grids of observers \obsA\ and \obsB.  The reasoning
  that led to this final representation is contained in the text of
  section 3.  An event with known coordinates in \obsA's frame, such
  as the event marked P, can now be represented in terms of \obsB's
  \xprime\ and \ctprime\ coordinates using the projection indicated by
  the dashed lines and then applying the scaling factors given in the
  text.}
\label{relativistic}
\end{figure}
Since the $y$ and $z$ transformations are trivial, the non-trivial
content of Lorentz transformations can be captured in the \xct\ graph
of Fig.~\ref{relativistic}.  In it, we are now able to correctly
represent observer \obsB's spacetime diagram on top of
observer \obsA's.  In Fig.~\ref{relativistic}, the grid lines parallel
to the $x$ and $x^\prime$ axes should be viewed as lines of
simultaneity for observers \obsA\ and \obsB\ respectively.  The lines
parallel to the $ct$ and $ct^\prime$ axes represent fixed positions in
their respective reference frames.

In order to make quantitative use of diagrams such as
Fig.~\ref{relativistic}, we need to know one more geometric quantity:
the size of unit intervals along the \xprime\ and \ctprime\ axes.
These are the intervals marked by stars and crosses on those axes.
They obviously have unit length in observer \obsB's own reference
frame, but get stretched in the Lorentz transformation $L_{{\bf B} \to
{\bf A}}.$ In Fig.~\ref{good_cells}, we defined $\gamma$ as the projection of the
unit \xprime\ interval onto the $x$ axis and the unit \ctprime\
interval onto the $ct$ axis.  From this and the slopes of the \xprime\
and \ctprime\ axes in Fig.~\ref{relativistic}, we find that
\begin{eqnarray}
\mathrm{length}(x^\prime ~ \mathrm{interval})
= \mathrm{length}(c t^\prime ~ \mathrm{interval})
= \gamma \sqrt{1 + v^2/c^2}
\end{eqnarray} 

Therefore, when converting a line-segment length along the \xprime\
axis into and an actual length that observer \obsB\ would measure, we
must divide by the factor $\gamma \sqrt{1 + v^2/c^2}.$ Similarly, when
converting a line-segment length along the \ctprime\ axis into an
actual time as judged by \obsB, we must divide by $c \gamma \sqrt{1 +
v^2/c^2}.$ We note that, a Lorentz transformation from \obsA's frame
to \obsB's frame is completely equivalent to (1) projecting an event,
such as event P in Fig.~\ref{relativistic}, onto the \xprime\ and \ctprime\ axes, as
shown by the dashed lines, and (2) converting the projected
line-segment lengths into length and time intervals, using the above
factors.

We find that it is usually more convenient to perform {\em Lorentz
transformations} algebraically using
Eqs.~\ref{lorentz_transformations}, which were derived here using
diagrams.  However, for many of the other results and applications of
special relativity, we believe that the method of Minkowski
diagrams is as illuminating or more illuminating compared to standard
algebraic methods.  Therefore, we point out these uses in
sections \ref{Results} and \ref{Paradoxes}.

Finally, the correct recipe for drawing observer \obsB's spacetime
diagram on top of observer \obsA's, as in Fig~\ref{relativistic}, is
the following:
\begin{enumerate}
\item
Draw the \ctprime\ axis as a line of slope $c/v$ that passes through
the origin of the \xct\ grid of observer \obsA.
\item
Draw the \xprime\ axis as a line of slope $v/c$ that passes through
the origin of the \xct\ grid of observer \obsA.  Thus, the angle
between the $x$ and \xprime\ axes is the same as the angle between the
$ct$ and \ctprime\ axes.
\item
Mark stars (clock ticks) along the \ctprime\ axis and crosses (unit length intervals) on
the \xprime\ axis, separated by a distance $\gamma \sqrt{1 + v^2/c^2}.$
\end{enumerate}

In our case, the above steps are motivated by the entire development
up to this point.  However, if one arrives at Lorentz transformations
through a different route, these three steps can be directly tied to
the Lorentz transformation equations
(Eqs.~\ref{lorentz_transformations}) as follows.  Since the
\ctprime\ axis is the $x^\prime = 0$ line, its equation on the
\xct\ plane can be found by setting to zero the l.h.s.\ of the first
equation in Eqs.~\ref{lorentz_transformations}.  This motivates step 1
above.  Similarly, step 2 can be motivated by setting $t^\prime = 0$
in the last of Eqs.~\ref{lorentz_transformations}.  Next, one can find
the $x^\prime = 1$ line and its intersection with the \xprime\ axis,
using Eqs.~\ref{lorentz_transformations}, to derive the size of unit
length intervals on the \xprime\ axis (step 3).  Unit
\ctprime\ intervals can be found in a similar way.

\subsection{Spacetime and the invariant interval}
\label{Spacetime}

In many introductory treatments, the idea of unifying space and time
into one entity---spacetime---is rationalized as follows: (1) unlike
with Galilean transformations where time is absolute, space and time
are ``mixed'' thoroughly by Lorentz transformations; (2) there exists
a metric for spacetime that remains invariant between inertial
observers.  We believe that, even at the introductory level, the
concept of spacetime needs to be justified with more quantitative
substance than this.  In particular, it should be pointed out that
{\em quantities of spacetime} are conserved by Lorentz
transformations.  What we mean here by a {\em quantity} or {\em
  amount} of spacetime is the 4-dimensional hyper-volume occupied by a
region of spacetime.

Algebraically, the above property is a consequence of {\em proper}
Lorentz transformations having unit determinant.  However, at the
introductory level this property can most easily be demonstrated
graphically.  Because the $y$ and $z$ transformations are trivial, a
2-d diagram such as Fig.~\ref{relativistic} is sufficient for our purposes.  In it, a
4-d hyper-volume of spacetime translates to a 2-d area on the \xct\
plane.

We start by showing that observer \obsB's unit cell---the highlighted
parallelogram of Fig.~\ref{relativistic}---has unit area, not just in
his own \xctprime\ plane, but also in observer \obsA's \xct\ plane.
If we denote the unit vectors along the $ct$ and $x$ axes to be
$\hat{l}$ and $\hat{i}$ respectively, the unit \xprime\ interval is
described by the vector $\gamma \hat{\imath} + \gamma (v/c) \hat{l};$
the unit \ctprime\ interval is described by $\gamma (v/c) \hat{\imath}
+ \gamma \hat{l}.$ The area of the parallelogram, equal to the
magnitude of the cross product of these vectors, is given by the
determinant
\begin{eqnarray}
\left| \begin{array}{cc}
\gamma & \gamma v/c \\
\gamma v/c & \gamma
\end{array} \right|
= 1.
\end{eqnarray}
Note that this is also the determinant of the Lorentz transformation
matrices $L_{{\bf A} \to {\bf B}}$ and $L_{{\bf B} \to {\bf A}}$ (see
Eqs.~(2) and (3)).

Now, suppose that one inertial observer, say \obsA, marks out a
certain patch of space-time on his position-time graph.  The number of
\obsB's unit cell parallelograms that would fit within this region is
equal to the number of \obsA's own unit cells that would fit within
it.  Thus, \obsA\ and \obsB\ would agree about the {\em amount} of
space-time that was marked out even though the shape of this patch
would look different in their own spacetime diagrams.  Therefore,
quantities/amounts of spacetime have physical meaning across inertial
reference frames.  This, we believe, strengthens the rationale for
adopting the concept of spacetime.

The invariance of the common metric used with spacetime---the
invariant interval---can be presented as a consequence of the above
result.  Given the 4-coordinates of two events, we can position
observer \obsA's coordinate system so that his origin coincides with
one of the events and the spatial separation between the two events
lies purely along the $x$ dimension.  Therefore, in \obsA's reference
frame, the separation between the events can be represented by the
vector $(\Delta x)\hat{\imath} + 0\hat{\jmath} + 0\hat{k} + c(\Delta
t) \hat{l},$ where $\Delta x$ and $\Delta t$ are the spatial and
temporal separations.  Next, we can construct a parallelogram on the
\xct\ plane using the above vector and its ``transpose,'' $c(\Delta
t)\hat{\imath} + (\Delta x)\hat{l}.$ Using the cross product of these
two vectors, one finds that the area of this parallelogram is $s^2 =
(\Delta x)^2 - c^2(\Delta t)^2.$ In observer \obsB's \xctprime\ plane
this parallelogram will deform into a different parallelogram with
area ${s^\prime}^2 = (\Delta x^\prime)^2 - c^2(\Delta t^\prime)^2.$
But we know that the two observers will agree on the area of this
parallelogram, which results in
\begin{eqnarray}
s^2 = {s^\prime}^2.
\end{eqnarray}

\section{Results and applications}
\label{Results}

The method of Minkowski diagrams is very useful for
understanding the standard set of ``results'' that arises from special
relativity.  These include time dilation, length contraction, velocity
addition, and the relativistic Doppler effect.  Most textbook problems
on special relativity deal with these results.  We stress that our
methods are equally well suited for introducing these topics as well
as for working problems related to them.

\subsection{Time dilation}
\label{Dilation}

With the above point in mind, we will discuss time dilation and length
contraction in terms of a problem that is typical of sophomore-level
textbooks: An observer on the ground sees an airplane traveling at
speed $v$ flying parallel to an airstrip of length $l$ on the ground.
Therefore, she finds that the airplane traverses the length of the
airstrip in time $l/v.$ (a) According to the pilot of the airplane,
how much time does the airplane take to traverse the length of the
airstrip?

\begin{figure}
\centering
\includegraphics[width=3in]{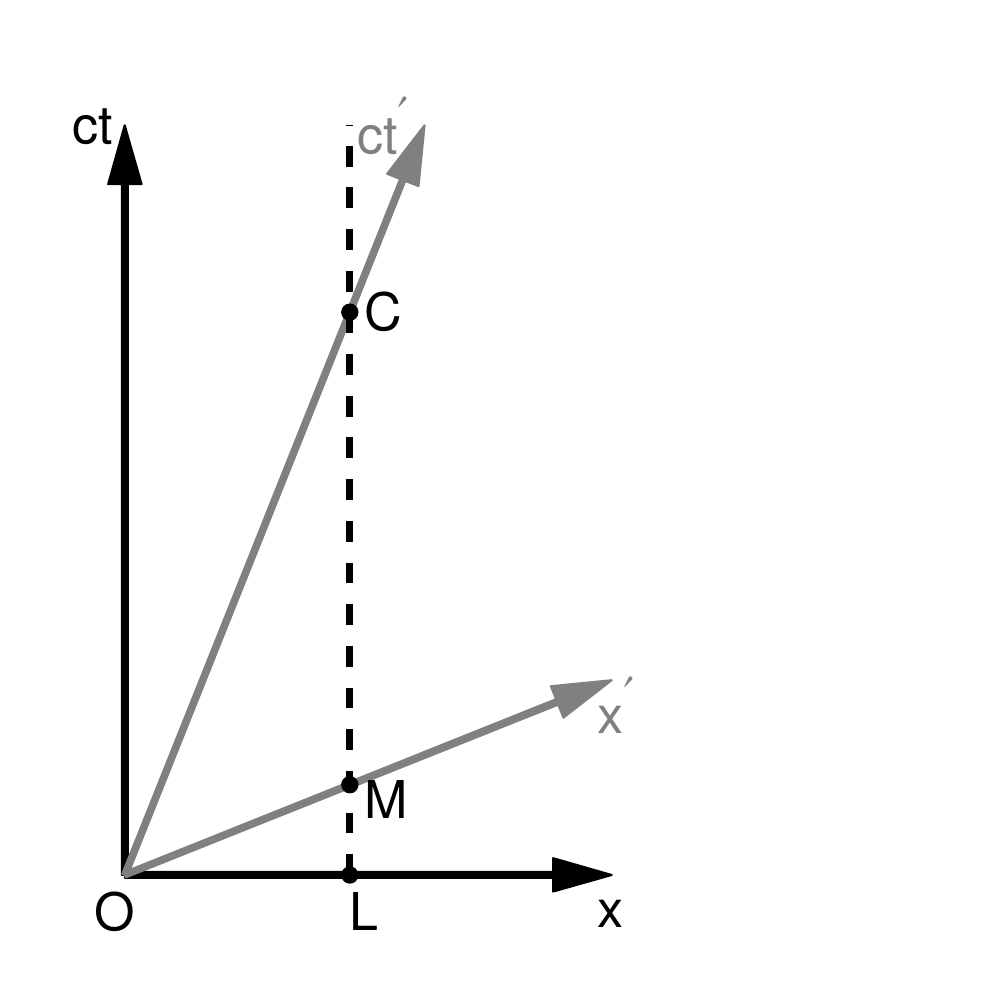}
\caption{The \xct\ frame belongs to the ground observer while the
  \xctprime\ frame belongs to the pilot (see text).  OL represents the
  airstrip.  The dashed line is the worldline of one end of the
  airstrip.}
\label{dilation}
\end{figure}
In Fig.~\ref{dilation}, the stationary frame (observer \obsA) is that
of the ground.  The axis labeled \ctprime\ represents the trajectory
of the front tip of the airplane and has slope $c/v.$ The line segment
OL represents the airstrip and has length $l.$ The worldline of the far
end of the airstrip (marked L) is represented by the vertical dashed
line.  The front tip of the airplane just reaching the far end of the
airstrip is represented by event C.  The time interval of interest is
the time recorded for event C in the two reference frames.  In the
ground frame, $c \Delta t$ is the length of LC, which can be computed
from the length of OL ($l$) and the slope of the \ctprime\ axis
($c/v$).  We find that $\Delta t = l/v,$ as stated in the problem.
The time of event C according to the pilot (observer \obsB) is
directly related to the length of line segment OC, which is found to
be $(cl/v) \sqrt{1 + v^2/c^2}$ from the Pythagorean theorem.  As noted
in section~\ref{Recipes}, this length must be divided by $ c \gamma
\sqrt{1 + v^2/c^2}$ in order to convert it to $\Delta t^\prime.$
Thus, the time measured by the pilot $\Delta t^\prime$ is different
from $\Delta t = l/v;$ it is $(l/v)/\gamma.$

Students often have difficulty deciding if a given time interval is a
{\it proper} time or not.  This is because they cannot visualize
events such as O and C---the front tip of the airplane intersecting
the beginning and then the end of the airstrip---occuring at the same
position in one observer's reference frame.  Fig.~\ref{dilation}
explicitly illustrates that OC is a proper time in the pilot's frame, as
it lies on the \ctprime\ axis ($x^\prime = 0$).  Given the generality
of the derivation, it will also be clear that ``improper'' time
intervals are always {\em dilated} by the factor $\gamma$ relative to
proper time intervals.

\subsection{Length contraction}

The natural continuation of the above problem is: (b) Does the pilot
of the airplane measure a different length for the airstrip, and if
so, what is that length?

Fig.~\ref{dilation} offers a good opportunity to stress what is meant
by {\em length.}  It is the instantaneous distance between the end
points of an object.  In other words, it is the size of an object
measured along a line of simultaneity.  For example, the unit length
intervals along the \xprime\ axis (marked with x's in
Fig.~\ref{relativistic}) are lengths for observer \obsB, while OL in
Fig.~\ref{dilation} represents a length for observer \obsA.  Thus, the
length of the airstrip as measured by the pilot is represented by the
line segment OM.  From the slope of the \xprime\ axis ($v/c$) and the
Pythagorean theorem, the length of OM is found to be $l \sqrt{1 +
  v^2/c^2}.$ As found in section \ref{Recipes}, this must be divided
by $\gamma \sqrt{1 + v^2/c^2}$ to yield a distance as measured by the
pilot.  Thus, the pilot measures the airstrip to have length
$l/\gamma.$ Given the generality of this derivation, it is clear that
the length of an object is largest in its own rest frame; it is
smaller by a factor of $\gamma$ when measured from a different frame.

\subsection{Velocity addition}
\label{Velocity}

\begin{figure}
\centering
\includegraphics[width=4in]{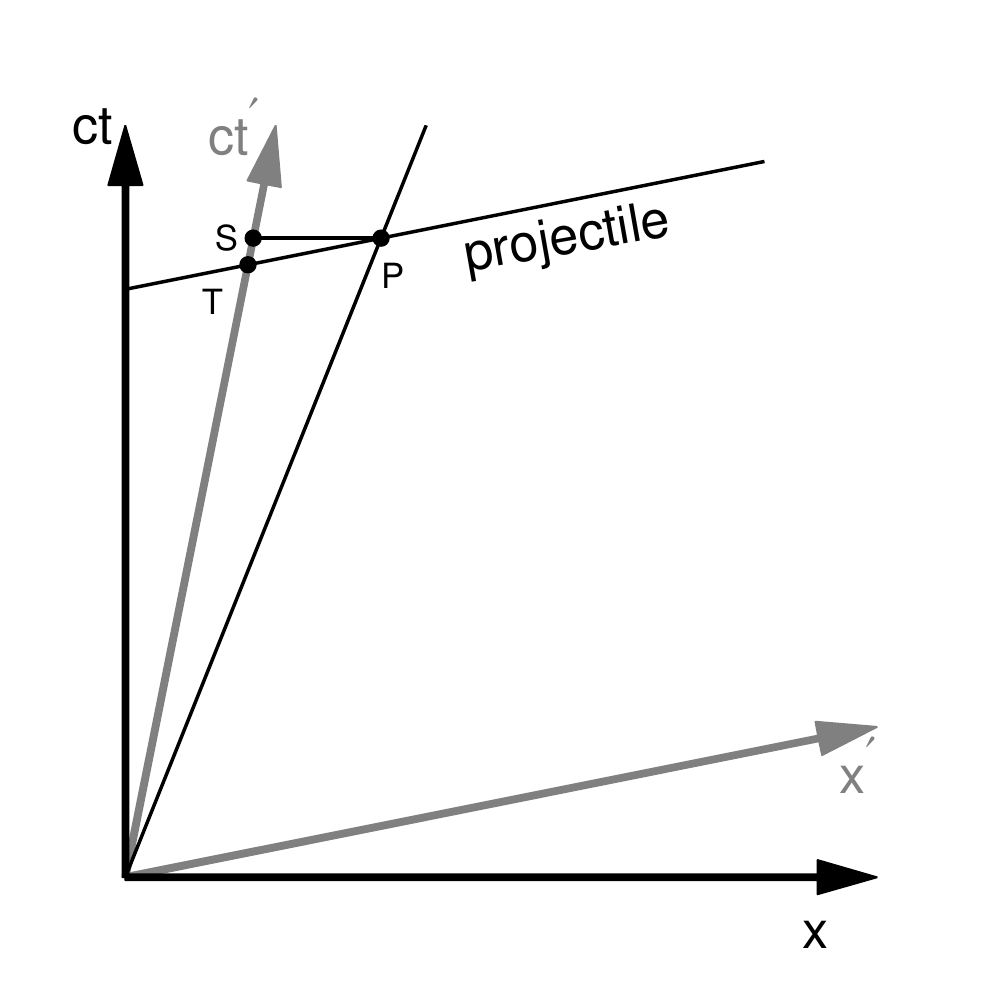}
\caption{Event T represents observer \obsB's clock registering 1 unit
  of time.  The tilted line (parallel to the \xprime\ axis) that goes
  through T is \obsB's unit-time line of simultaneity.  It intersects
  the projectile's worldline at point P.}
\label{velocity}
\end{figure}

The question addressed here is: if a projectile moves with constant
velocity $\vec{U} = U_x \hat{\imath} + U_y \hat{\jmath} + U_z \hat{k}$
in observer \obsA's reference frame, what is its velocity
$\vec{U^\prime} = U_x^\prime \hat{\imath} + U_y^\prime \hat{\jmath} +
U_z^\prime \hat{k}$ in \obsB's reference frame?  A projection of this
situation along the $x$ direction is shown in Fig.~\ref{velocity}.
One way for \obsB\ to establish the projectile's velocity is to
measure the $x^\prime,$ $y^\prime,$ and $z^\prime$ coordinates of the
projectile when exactly 1 time unit has elapsed on his own clock.

In Fig.~\ref{velocity}, event T indicates \obsB's clock registering 1
unit of time.  We know that events simultaneous with this time are
identified by the light solid line parallel to the \xprime\ axis.
From the section on time dilation above (section \ref{Dilation}), we
know that this line intersects the $ct$ axis at $c(1/\gamma).$
Therefore, the expression for observer \obsB's unit-time simultaneity
line is $ct = (v/c)x + c/\gamma.$ The projectile's motion is described
by the expression $ct = (c/U_x)x.$ Therefore, by setting these two
equal, the coordinates of the event marked P are found to be
\begin{eqnarray}
x_\mathrm{P} = {U_x \over \gamma(1 - U_x v / c^2)}\\
t_\mathrm{P} = {1 \over \gamma(1 - U_x v / c^2)}.
\label{xPtP}
\end{eqnarray}
Now, using Eq.~\ref{xPtP}, we can easily write down $y_\mathrm{P}$ and $z_\mathrm{P},$
the $y$ and $z$ coordinates of the projectile at time $t_\mathrm{P}.$
Since the Lorentz transformation for $y$ and $z$ are trivial (see
Eq.~\ref{lorentz_transformations}), we find that
\begin{eqnarray}
U_y^\prime = y_\mathrm{P}^\prime = y_\mathrm{P} = U_y t_\mathrm{P} =
{U_y \over \gamma(1 - U_x v / c^2)}\\
U_z^\prime = z_\mathrm{P}^\prime = z_\mathrm{P} = U_z t_\mathrm{P} =
{U_z \over \gamma(1 - U_x v / c^2)}.
\end{eqnarray}

A common problem with relativistic velocity addition is the difficulty
in intuiting why $U_x$ appears in the expressions for $U_y^\prime$ and
$U_z^\prime,$ especially since the Lorentz transformations for $y$ and
$z$ and trivial.  Fig.~\ref{velocity} is very useful in this regard.
The critical step is to understand the physical significance of event
P.  It represents the coordinates of the projectile, as judged by
observer \obsB, exactly when his clock registers 1.  Therefore,
$t_\mathrm{P}^\prime = 1.$ However, $t_\mathrm{P}$ clearly depends on
the slope of the projectile's trajectory on the \xct\ plane, and hence
on $U_x.$ It is through $t_\mathrm{P}$ that $U_x$ finds its way into
the expressions for $U_y^\prime$ and $U_z^\prime.$

Next, we need to find $U_x^\prime.$ This is equal to
$x_\mathrm{P}^\prime,$ which is represented by line segment TP in
Fig.~\ref{velocity}.  Knowing the equations for all the lines
involved, it is possible to obtain the length of TP algebraically.  On
the other hand, we can make quick progress by letting TP represent a
{\em ruler} in \obsB's reference frame.  Note that the line segment SP
then represents the length of this ruler as determined by
\obsA.\cite{TPruler} But
\begin{eqnarray}
{x_\mathrm{P}^\prime \over \gamma}
= \mathrm{length(SP)} = (U_x - v) t_\mathrm{P},
\end{eqnarray}
where the first equality follows from Lorentz-contraction and the
second equality is simply obtained from the relative speeds of the
projectile and observer \obsB.  Substituting $t_\mathrm{P}$ from
Eq.~\ref{xPtP}, we find that
\begin{eqnarray}
U_x^\prime = x_\mathrm{P}^\prime =
{U_x - v \over 1 - U_x v / c^2}.
\end{eqnarray}

\subsection{Relativistic Doppler effect}

It is in regard to the relativistic Doppler effect that
Minkowski-diagram-based methods have reappeared in this
journal.\cite{Reynolds1990Cook1991} For completeness, we will quickly
demonstrate why this method is ideal for deriving the Doppler
equations.  In our example, signals travel at speed $v_s,$ which is
not necessarily equal to $c.$ The results obtained converge to the
Doppler equations for light when $v_s$ is set to $c.$

\begin{figure}
\centering
\includegraphics[width=3in]{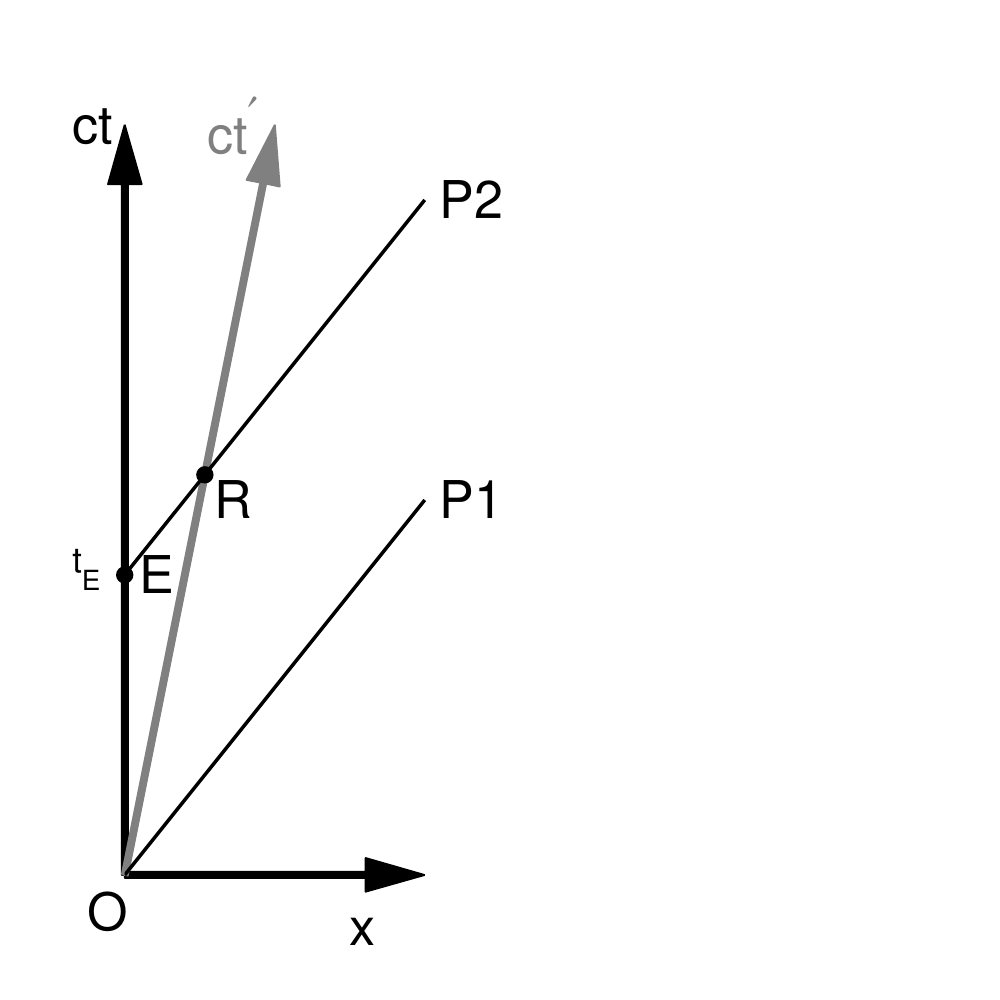}
\caption{Pulses, designated P1, P2 etc., are emitted at a period $t_E$
  in observer \obsA's reference frame.  Their reception by observer
  \obsB\ is represented by the intersection of these pulses with the
  \ctprime\ axis.}
\label{doppler}
\end{figure}
Fig.~\ref{doppler} indicates that observer \obsA\ emits pulses (or
wave crests) with a period $t_\mathrm{E}$ starting at time $t = 0.$
The first pulse, labeled P1 in Fig.~\ref{doppler}, is received by a
{\em receding} observer (observer \obsB) at $t^\prime = 0.$ The
question is: at what time $t_\mathrm{R}^\prime$ does observer
\obsB\ receive P2, the second pulse?  We can express the worldline of the
second pulse as $ct = (c/v_s)x + c t_\mathrm{E}.$ As usual, the
\ctprime\ axis may be expressed as $ct = (c/v)x.$ By equating these
two expressions, the coordinates of event R (the reception of P2 by
observer \obsB) are found to be
\begin{eqnarray}
x_\mathrm{R} = {t_\mathrm{E} \over 1/v - 1/v_s} \\
t_\mathrm{R} = {t_\mathrm{E} \over 1 - v/v_s}.
\end{eqnarray}
Thus,
\begin{eqnarray}
\mathrm{length(OR)} =
\sqrt{c^2 t_\mathrm{R}^2 + x_\mathrm{R}^2} = {c t_\mathrm{E} \over 1 -
v/v_s} \sqrt{1 + v^2/c^2}.
\end{eqnarray}
As discussed in section \ref{Recipes}, $\mathrm{length(OR)}$ must be
divided to $c \gamma \sqrt{1 + v^2/c^2}$ to obtain
$t_\mathrm{R}^\prime.$ Thus, according to \obsB, the time of reception
of the second pulse
\begin{eqnarray}
t_\mathrm{R}^\prime = {t_\mathrm{E} \over \gamma (1 - v/v_s)}.
\end{eqnarray}
So, the Doppler shift in frequency may be written as
\begin{eqnarray}
{f^\prime \over f} = {t_\mathrm{E} \over t_\mathrm{R}^\prime}
= \gamma (1 - v/v_s).
\end{eqnarray}
The case of {\em approaching} emitter and receiver may be treated in a
very similar way after extending Fig.~\ref{doppler} into the lower
left quadrant where both $x$ and $ct$ are negative.

\subsection{Other textbook problems}

There is a class of textbook problems that calls for Lorentz
transformations from one reference frame to another.  As mentioned in
section \ref{Recipes}, an event with known coordinates in observer
\obsA's reference frame can be projected onto observer \obsB's
reference frame using lines parallel to the \xprime\ and
\ctprime\ axes.  Thereafter, it {\em is} possible to find the
\xprime\ and \ctprime\ values of the event using purely geometry.
However, we find that geometrical solutions turn out to be as lengthy
as the derivation of the Lorentz transformation.  Therefore, we
recommend using Eqs.~\ref{lorentz_transformations} for these problems,
which were derived here using Minkowski diagrams.

In another class of problems, the 4-coordinates of two spatially and
temporally separated events are provided.  Students are asked whether
these two events can occur at the same position in some observer's
reference frame and, if so, what the velocity of that observer needs
to be.  Alternately, they are asked if the two events can be
simultaneous for some observer and, if so, what the velocity of the
that observer needs to be.  These problems can be solved very easily
using Minkowski diagrams and, in our experience, most relativity
instructors/texts do recommend spacetime diagrams for solving problems
of this sort.

\section{Paradoxes}
\label{Paradoxes}

The major theme that links the famous relativistic paradoxes is the
relativity of simultaneity.  The graphical methods developed here are
especially helpful for visualizing the latter and, therefore, for
unraveling paradoxes.  We will briefly describe how four well
known paradoxes can easily be visualized using the methods developed
here.

\subsection{The Andromeda paradox}

In the following, we will refer to an observer's lines of simultaneity
as his/her ``time-frames.''  From Minkowski diagrams, it is clear that
even if observer \obsB\ moves at a walking pace relative to \obsA, his
time-frames are very slightly tilted relative to \obsA's time-frames.
According to Fig.~\ref{relativistic}, events in the positive $x$
direction, that are in \obsB's {\em present} time-frame are in a {\em
  future} time-frame of observer \obsA.  Events in \obsB's present
time-frame that are in the negative $x$ direction are in the past of
observer \obsA.  Even when $v$ (the relative velocity between
\obsA\ and \obsB) is very small, a difference in the time assigned to
a remote event by \obsA\ and \obsB\ can be quite large when the event
is located very far away.  For instance, suppose that intelligent
beings within the Andromeda galaxy have just now learned of the
existence of humans on earth.  Now, by starting to walk in the
direction of the Andromeda galaxy you will be able to ``dial in'' to
your time-frame a later date on the Andromedan's calendar.  Perhaps by
that date, after careful consideration, they have already launched a
fleet of spaceships to conquer the earth.

It is important to stress that this time-frame jump is a non-local (or
remote) effect and that, for instance, we cannot use it to influence
the eventual actions of the Andromedan fleet.  However, as we will see
in the context of the next paradox, the ability to dial in various
dates on a remote calendar into one's time-frame can lead to
interesting local effects as well.

\subsection{The twin paradox}

In this well known paradox, one twin stays on the earth while the
other races to a distant point in space at a significant fraction of
$c,$ quickly turns around, and returns to earth at the same high
speed.  When they reunite, the earth-bound twin has aged more than the
astronaut twin.  The apparent paradox is that the motion of the
astronaut twin relative to the earth-bound twin is exactly the same as
the motion of the earth-bound twin relative to the astronaut twin.
Therefore, how can there be an asymmetry in their aging?

\begin{figure}
\centering
\includegraphics[height=2.5in]{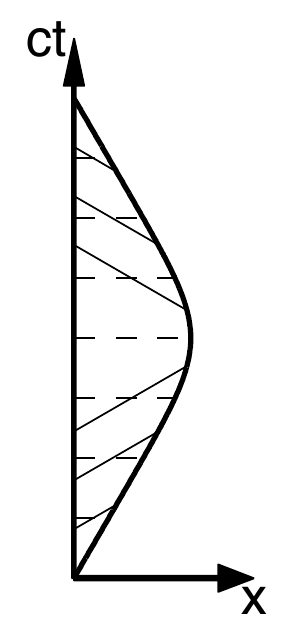}
\caption{The thick solid curve that moves away from the $ct$ axis and
then returns to it is the astronaut twin's worldline.  The dashed
lines and the light solid lines are lines of simultaneity of the
earth-bound and the astronaut twins respectively.}
\label{twins}
\end{figure}

In most textbooks, it is explained that the simple time dilation
calculation (dividing by $\gamma$) is only applicable from an
inertial, or non-accelerating, frame.  Thus, we can employ the simple
calculation from the reference frame of the earth-bound twin but not
the astronaut.  Therefore, the answer obtained this way by the
earth-bound twin---that the astronaut twin ages less---must be
correct.  To strengthen this argument, many authors use spacetime
diagrams and follow the transmission and reception of light signals
issued by the two twins.  These diagrams are very good at dispelling
any doubts that the astronaut twin ages less.  However, these
approaches fall short of pinpointing, as the explicit cause of unequal
aging, the intense acceleration of the astronaut twin during the
turn-around.  With Minkowski diagrams, we can directly visualize how
one twin perceives the passage of time of the other twin, as
illustrated in Fig.~\ref{twins}.

The dashed horizontal lines are the earth-bound twin's time-frames
(lines of simultaneity) that mark one-year intervals since the
astronaut's departure.  The tilted solid lines are the astronaut's
time-frames, separated along her worldline by intervals that
correspond to a year.  During most of the out-bound and return legs of
the trip, both twins observe the other's time progressing at a fixed
but slow rate.  For instance, the earth-bound twin's year-2 time-frame
(the second dashed horizontal line from the bottom) intersects the
astronaut's worldline before year-2 arrives on the astronaut's
calendar.  Similarly, the astronaut's year-2 time-frame intersects the
$ct$ axis below the second dashed line.  Note that the astronaut's
time-frames have a slope of $v/c,$ where $v$ is the speed of the
astronaut at the intersection of the astronaut's time-frame and
worldline.  Therefore, between year 3 and year 4 the slope of these
time-frames undergoes a rapid change (from positive to negative).  As
a result, she ``dials in'' a quickly incrementing sequence of dates on
the earth-bound twin's calendar into her own time-frame.  Thus, from
her point of view, the earth-bound twin ages very quickly during the
turn-around.  In fact, at the end of year-4 on the astronaut's
calendar, she estimates that the earth-bound twin has aged $> 5$
years.  Although the mutual perception that the other twin ages more
slowly is restored soon after the turn-around, the astronaut twin can
never catch up with this difference and returns to earth one year
younger than the earth-bound twin.  On the other hand, the earth-bound
twin ``judges'' that the astronaut ages at essentially the same slow
rate throughout the journey.\cite{TwinNote}

\subsection{The pole vaulter paradox}

A group of physicists convince a pole vaulter to run very fast while
carrying his pole, in order to demonstrate Lorentz contraction.  To
prove their point, the physicists ask the pole vaulter to run through
a barn that is shorter than the rest-frame length of the pole.  At the
beginning of the demonstration, the {\em entry door} of the barn is
open and the {\em exit door} of the barn is closed.  According to the
physicists' plan, the pole vaulter will enter the barn through the
entry door.  Once the contracted pole is fully contained within the
barn, the entry door will be closed first.  Next, the exit door is
opened to allow pole vaulter to exit the barn without accident.  Of
course, the physicists at each door must synchronize their clocks and
be able to open/close doors by a sufficient amount almost
instantaneously.  Even so, if the pole vaulter knew about relativity,
he would not have agreed to this exercise, would he?  For, in his
frame, the barn would appear even shorter than normal compared to the
pole, due to Lorentz contraction.  Thus, the pole would never fit
within the barn and would surely collide with at least one of the
doors because there is never a time when both doors are open.  The
physicists, on the other hand, believe that no collision will take
place.  In the end, only one answer---collision or no collision---can
be correct.  But none of the viewpoints expressed above seem to be
wrong.

\begin{figure}
\centering
\includegraphics[width=2.5in]{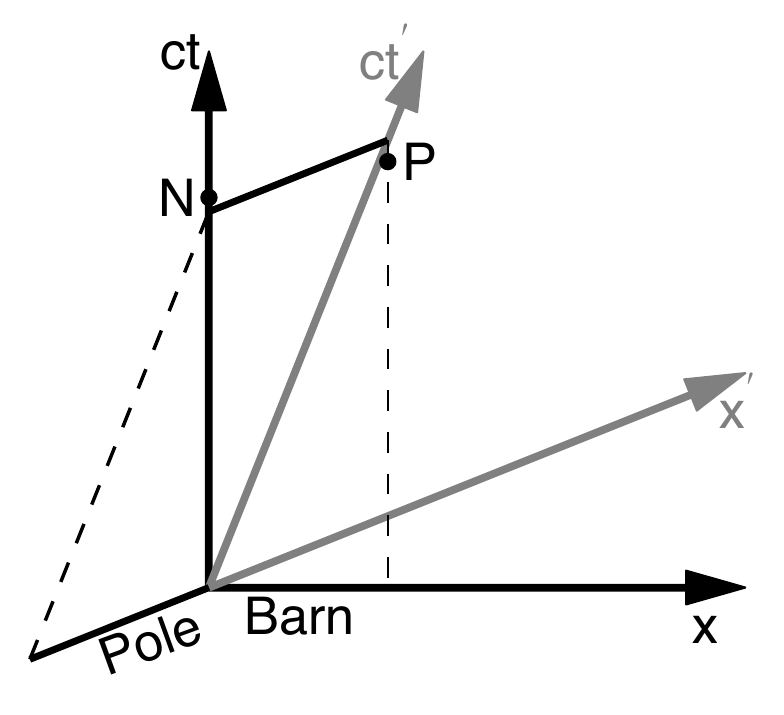}
\caption{The region between the $ct$ axis and the vertical dashed line
  is the spacetime traced out by the barn.  The tilted dark solid
  lines are snapshots of the pole in its rest frame at two points in
  time.  The tilted dashed line and the \ctprime\ axis are the
  worldlines of the two ends of the pole.  Event N represents the
  closing of the entry door and event P represents the opening of the
  exit door.}
\label{pole}
\end{figure}
The key to this paradox is the difference between the pole vaulter's
and physicists' time-frames.  Fig.~\ref{pole} illustrates the
resolution of this paradox quite simply.  The physicists frame is the
stationary one (observer \obsA).  The barn and the pole are labeled
along the $t = 0$ and $t^\prime = 0$ time-frames respectively.  The
two critical events in this demonstration---the closing of the entry
door and the opening of the exit door---are events N and P
respectively.  The key here is to realize that the pole vaulter's
lines of simultaneity are parallel to the pole in Fig.~\ref{pole} and,
therefore, events N and P occur in reverse order.  Thus, in the pole
vaulter's reference frame, the exit door opens {\em before} the entry
door closes!  Thus, the pole vaulter must agree with the physicists
that no collision will take place.

\subsection{Bell's paradox}

Two rockets in space are at rest relative to the earth.  They are tied
together with a stretched rope.  If the rope is stretched anymore, it
will break.  The pilots of the rockets set off on a journey but agree
to reach cruising speed by using a synchronized sequence of
accelerations.  The pilots believe that, because they accelerate in
unison, the rocket-rope-rocket system will reach cruising speed as one
object, and the rope will not be stretched anymore.  Therefore, it
will not break.  The pilots point out that those on earth would
observe the gap between the two rockets (occupied by the rope) to
Lorentz contract.  On the other hand, an earth observer insists that
he would measure the gap between the rockets to remain unchanged if
the pilots do indeed accelerate their rockets in a synchronized
manner.  He argues that the pilots, in their frame, would find the
separation between rockets to increase.  Therefore, the rope would
break.  The object here is to determine whose argument is flawed and,
therefore, whether the rope breaks or not.

\begin{figure}
\centering
\includegraphics[width=3in]{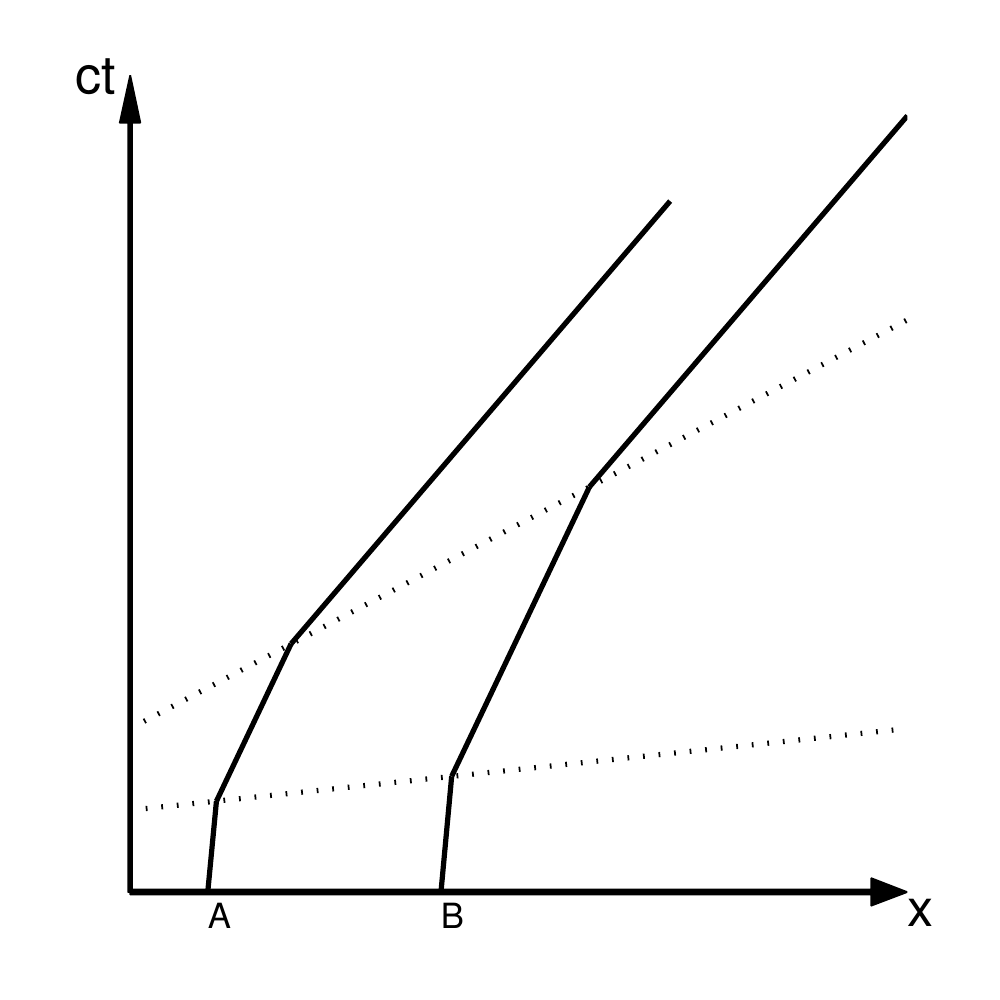}
\caption{A and B are two rockets initially at
  rest in the earth's frame.  After the first thrust puts the rockets
  in motion, their time-frames are tilted and the second thrust is
  synchronized along one such time frame, indicated by the lower
  dotted line.  The third thrust is synchronized along a further tilted
  time frame, indicated by the upper dotted line.}
\label{rockets}
\end{figure}
It turns out that neither argument is wrong, just that the statement
``synchronized accelerations'' cannot apply to both the pilots'
frame (once they start moving) and the earth frame.  If the rocket
thrusts appear to be synchronized to the earth observer, the
pilots' would detect a delay between the thrusts of the two
rockets.  In that case, the assertion of the earth observer, that the
rope will break, will come true.  In Fig.~\ref{rockets}, we illustrate
the opposite case where the thrusts are synchronized in the pilots'
frame.  The three thrusts lie along the pilots' lines of
simultaneity at the time.  In the earth frame (the stationary one), rocket B
always accelerates later, causing the gap between the two to contract.
Therefore, in this case, the rope does not break.

\section{Synopsis}
\label{Synopsis}

Minkowski diagrams deserve a prominent place among the pedagogical
tools used for introducing special relativity to undergraduates.  The
graphical lessons presented here can serve as the primary means of
instruction or an alternate route available to students.  The major
advantages of the methods presented here are:
\begin{enumerate}
\item
The use of diagrams helps students visualize situations and
facilitates qualitative reasoning.  This is key to true learning and
retention.  The availability of graphical alternatives is important
for enabling a more conceptual and intuitive grasp of relativity as
opposed to math-based proficiency.
\item
Converting qualitative reasoning to quantitative answers requires just
a few simple steps, as outlined in section \ref{Recipes}.   The
brevity of this section illustrates how easily these diagrams can be
adapted for quantitative use.
\item
The usefulness of Minkowski diagrams is not limited to a subset of
introductory topics.  Methods based on them are as effective and
elegant as any other treatment of the complete cannon of introductory
topics, that range from Galilean transformations and the notion of
inertial observers, to deriving Lorentz transformations, time
dilation, length contraction, and other important results, to
resolving difficult paradoxes in relativity.
\end{enumerate}
Our main goal has been to present, in one place, a complete set of
introductory lesson plans based on Minkowski diagrams.  We have also
pursued a modern and systematic approach suited to present-day
undergraduate instruction.

\appendix*

\section*{Appendix: Alternate method for finding the value of $\gamma$}

\begin{figure}
\centering
\includegraphics[width=2.5in]{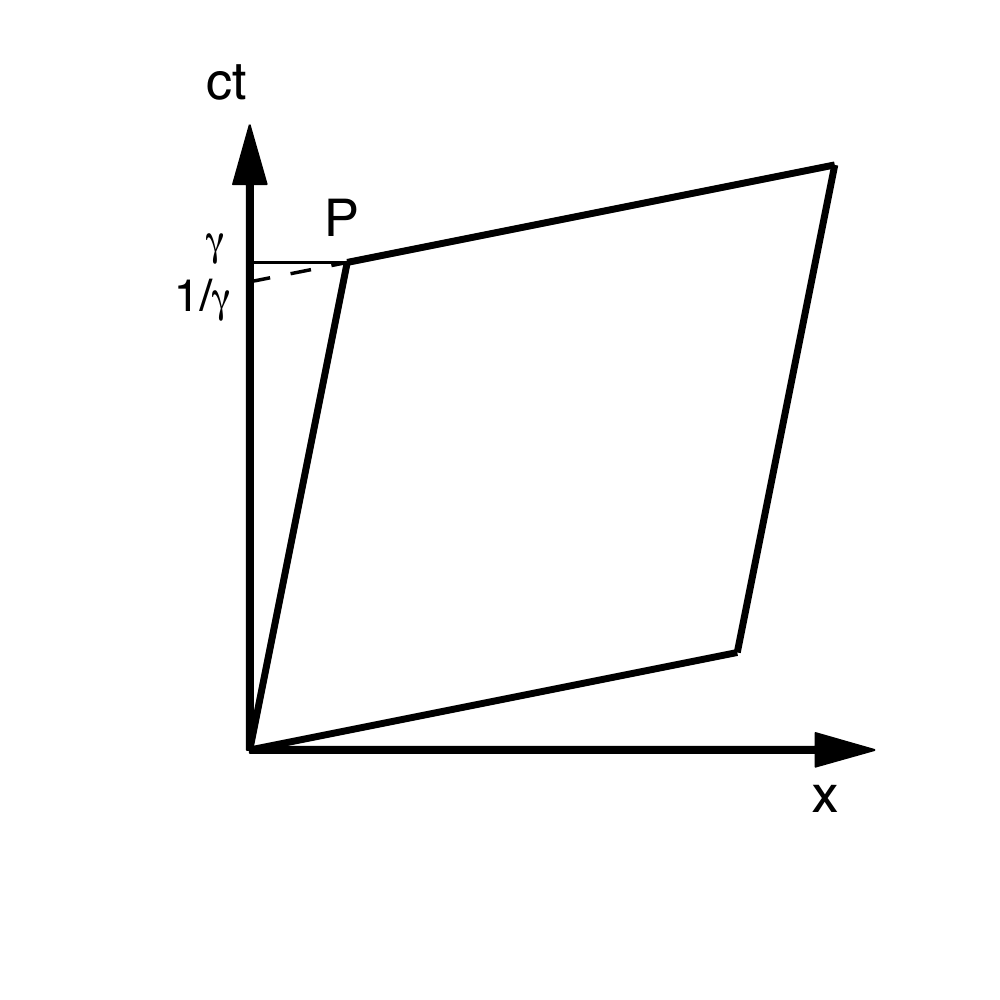}
\caption{The parallelogram is the unit cell of observer \obsB's
  position-time grid.  Event P is \obsB's first clock tick ($ct^\prime
  = 1$). The thin solid line is \obsA's line of simultaneity through this
  event.  The dashed line identifies all events that are simultaneous
  with P according to observer \obsB.}
\label{gamma_geom}
\end{figure}

Here, we find the value of $\gamma$ using a purely geometrical
approach as an alternative to the derivation of section \ref{Lorentz}.
Note that $\gamma,$ at this point, is simply a geometrical factor that
sets the size of the unit parallelogram of observer \obsB, as shown in
Fig.~\ref{gamma_geom}.
The projection of event P, the top left corner of \obsB's unit cell,
onto the $ct$ axis has the following physical meaning.  Observer
\obsA\ judges that the first tick of observer \obsB's clock (when
$ct^\prime = 1$) is simultaneous with his own clock reading a value of
$ct = \gamma,$ not 1.  Observer \obsB\ should come to the same
conclusion about the rate at which observer \obsA's clock advances.
Therefore, according to him, \obsA's clock reads $ct = 1/\gamma$ when
his own clock reads $ct^\prime = 1.$ The dashed line of
Fig.~\ref{gamma_geom} represents \obsB's unit-time line of
simultaneity.  Since the $ct$ intercept of this line is $1/\gamma$ and
the slope of \obsB's line of simultaneity is $v/c,$ the expression
for the dashed line is $ct = (v/c)x + 1/\gamma.$  The \ctprime\ axis
may be described by the expression $ct = (c/v)x.$ By equating these
two expressions, we find the \xct\ coordinates of event P to be
\begin{eqnarray}
x_\mathrm{P} = {v/c \over \gamma (1 - v^2/c^2)} \\
ct_\mathrm{P} = {1 \over \gamma (1 - v^2/c^2)}.
\label{ctP}
\end{eqnarray}
But $\gamma$ is defined as $ct_\mathrm{P}$ in Fig.~\ref{gamma_geom}.
Therefore, it follows from Eq.~\ref{ctP} that $\gamma^2 = 1/(1 -
v^2/c^2).$

\begin{acknowledgments}

We are indebted to Monica Moore (formerly the science division liaison
of the Illinois Wesleyan University Library and now at Notre Dame)
for her untiring efforts in researching the use of Minkowski diagrams
in the literature.  Her searches extended well beyond the the period
for which electronic abstracts are available.

\end{acknowledgments}


\begin{thebibliography}{10}
\bibitem{ShadowitzGRnote}
Shadowitz also notes that, in comparison to other graphical approaches, this
  one is the most convenient for extension to 4 dimensions and, therefore, for
  general relativistic applications.
\bibitem{GammaGeom}
We want to emphasize here that $\gamma$ at this point has no physical meaning.
  It is simply a geometrical scaling factor.
\bibitem{TPruler}
This is true because the \ctprime\ axis is also the worldline of the left
end of the ruler TP.
\bibitem{TwinNote}
Actually, according to the earth-bound twin, the astronaut twin ages slightly
  faster near the turn-around, because the slope of her worldline is changing.
  For instance, when the astronaut's worldline is vertical, both twins age at
  the same rate.
\end{thebibliography}

\end{document}